\documentclass[a4paper,fleqn,usenatbib]{mnras}
\usepackage[T1]{fontenc}
\usepackage{ae,aecompl}
\usepackage{amsmath}
\usepackage{amsfonts}
\usepackage{amssymb}
\usepackage{graphicx}
\usepackage{color,xcolor}
\usepackage{orcidlink}
\def\mnras{MNRAS}
\def\aap{A$\&$A}

\def\apj{ApJ}
\def\apjl{ApJ Letters}
\def\apjs{ApJS}
\def\pasp{PASP}
\def\aj{AJ}

\newcommand{\me}{\, {\rm M}_{\oplus}}
\newcommand{\msun}{\, {\rm M}_{\odot}}

\newcommand{\au}{\, {\rm au}}

\newcommand{\sigori}{$\sigma~{\rm Orionis}$}
\newcommand{\sigoristar}{$\sigma~{\rm Ori~AB}$}
\title[$\sigma$ Ori the runaway]{Runaway origins of a disc mass gradient in $\sigma$ Orionis}
\author[G. A. L. Coleman et al]{Gavin A. L. Coleman\orcidlink{0000-0001-5111-8963}$^{1}$\thanks{Email: gavin.coleman@qmul.ac.uk}, Thomas J. Haworth\orcidlink{0000-0002-9593-7618}$^{1}$, and Jinyoung Serena Kim\orcidlink{0000-0001-6072-9344}$^{,2}$\\
$^1$Astronomy Unit, Department of Physics and Astronomy, Queen Mary University of London, Mile End Road, London, E1 4NS, UK\\
$^2$Steward Observatory, University of Arizona, 933 N. Cherry Ave, Tucson, AZ 85721-0065, USA\\
}
\date{Accepted 2025 September 24. Received 2025 September 4; in original form 2025 July 16}
\pubyear{2025}
\begin{document}
\label{firstpage}
\pagerange{\pageref{firstpage}--\pageref{lastpage}}
\maketitle
\begin{abstract}
Radiation from massive stars is known to significantly affect the evolution of protoplanetary discs around surrounding stars by driving ``external’’ photoevaporative winds. Typically most studies assume that the massive stars driving these winds are comoving with their associated clusters. However, it is also known that massive stars can be runaways, after being violently ejected from their birth environment through interactions with other massive stars. In this letter, we show that the well studied system \sigoristar~is actually a runaway system, only now passing through \sigori. There are multiple observable features that indicate this is the case, including significantly larger proper motions for \sigori~than the surrounding stars, an infrared arc of ionising gas along the predicted velocity vector, and a disparity in protoplanetary disc masses across \sigori. We finally use protoplanetary disc evolution models to explain the observed disparity in disc masses, showing that those discs downstream of \sigoristar, i.e. those yet to encounter it, have larger masses than those upstream, consistent with observations. Overall, our work highlights the importance of understanding the dynamical history of star forming regions, since the time varying UV fields provided by runway stars results in a complex history for the evolution of the protoplanetary discs.

\end{abstract}

\section{Introduction}
\label{sec:intro}

High energy UV radiation from massive stars in star forming regions is now known to disperse gas around nearby protoplanetary discs, through the launching of photoevaporative winds \citep{Winter23}. These ``external'' photoevaporative winds are often seen in observations as cometary-like features known as ``proplyds'' \citep{Odell94,Bally2000, Odell01, Ricci08, Kim16, Haworth21, PastPresentFuture2025}. 

The \sigori~cluster has been subject to study in the context of external photoevaporation. Though no imaging of proplyds has yet taken place, there are members of the cluster with excess optical emission attributed to external photoevaporation \citep{Rigliaco09,Natta14,Ballabio23, Mauco2025}.  Discs in \sigori~also display distributions in disc masses relative to their proximity to the massive star system in the region, \sigoristar\footnote{The system \sigoristar, is actually a triple star system including a close spectroscopic binary (Aa,Ab) with an additional companion separated by 0.25'' (B). All three stars have masses between 11.5--17 $\msun$ \citep{Schaefer16}}, where there is a depletion in disc masses observed as the projected distance decreases \citep{Ansdell17,Mauco23}. This observed depletion can be explained using external photoevaporation models \citep{Winter20b}, but these models typically assume that the cluster is relatively co-moving.

Within star forming regions, there is expected to be significant stellar dynamics, as newly formed stars and associations mutually interact. These interactions can result in stars being ejected from within clusters, typically referred to as walkaway or runaway stars \citep[for a recent catalogue see][]{CarreteroCastrillo25}. Indeed, $\sim20\%$ of OB stars end up as runaway stars early on in their lifetimes \citep[e.g.][]{Fujii11}. After these runaway stars are ejected from a cluster, they can interact with other star forming clusters, dispersing protoplanetary discs to similar levels as if they were co-moving \citep{Coleman25FUV}. For this reason, it means that whilst statistical imprints of external photoevaporation on disc properties have been observed \citep[e.g.][]{Ansdell17,Eisner18,VanTerwisga20,Mauco23}, they may be difficult to adequately interpret. Other processes also contribute to the difficulty in interpreting disc properties in stellar clusters, including shielding by the natal molecular cloud \citep{Qiao22,Wilhelm23}, and different formation times \citep{Coleman22}.

In this paper, we explore the possibility that \sigoristar~is a runaway system and is performing a flyby of \sigori. We examine: the differences in proper motions and radial velocities between \sigoristar~and the remainder of the cluster; observable impacts of the ionising winds from \sigoristar~on the surrounding gas; and observable differences in protoplanetary disc masses relative to the location of those discs in terms of the expected motion vector of \sigoristar. Finally, we ran numerical simulations of evolving protoplanetary discs to validate the distributions seen in observed disc masses, and then provide predictions for how observations of protoplanetary disc radii can confirm our interpretation that \sigoristar~is indeed a runaway system.

\section{Observational evidence that $\sigma$ Orionis may be a runaway}

\subsection{Large Proper Motions}

Previous studies of \sigori~typically presume that the stars are relatively co-moving and that the stars, and therefore their resident discs, have evolved in this way for the expected ages of the systems.
Indeed, the majority of stars do have similar proper motions averaging to $\mu_{\alpha} = 1.49 \rm mas~yr^{-1}$ and $\mu_{\delta} = -0.62 \rm mas~yr^{-1}$ \citep{Mauco23,Zerjal24}. Additionally, the radial velocities of the stars in \sigori\ were initially found to be $31\pm0.6\rm km~s^{-1}$ using pre \textit{Gaia} data \citep{Jeffries06}, whilst analysis of \textit{Gaia} data confirmed these velocities by finding an average of $30.7\rm km~s^{-1}$ \citep{Zerjal24}.
However, there is still some variance in those values across the cluster, most likely due to dynamical interactions early in its life.

The one exception to the small deviations in proper motions and radial velocities in \sigori, appears to be \sigoristar~itself. Indeed, \citet{VanLeeuwen07} reanalysed \textsc{hipparcos} data to determine the proper motions for \sigoristar, finding $\mu_{\alpha}=22.63\pm10.83$mas yr$^{-1}$ and $\mu_{\delta}=13.45\pm 5.09$mas yr$^{-1}$. It is worth noting that the \textsc{hipparcos} observations would have measured the photo-centre of $\sigma$ Ori Aa, Ab, and so would include systematic errors based on the orbital configurations at the times of exposure. However this error should be accounted for in the errors on the proper motions, given that the projected range of the orbit, the inner binary is much smaller than the calculated proper motion values, and that the astrometric data was taken over a 4 year period, thus including multiple orbits of the inner binary. Additionally, the effects of the outer companion on the proper motion calculations should be effectively unchanged over the time of the observations, given the orbital period of the outer binary, $\sim 160$ yrs \citep{Schaefer16}, is much longer than the \textsc{hipparcos} data baseline. In terms of the on-sky motion, these proper motions correspond to $\mu_{\alpha}\sim 40.7\rm km~s^{-1}$ and $\mu_{\delta}\sim 24.1\rm km~s^{-1}$, assuming a distance of 387.5 pc to the system as determined by \citet{Schaefer16}, a significant improvement on previous estimates. Whilst more recent reductions of \textit{Gaia} DR3 data has calculated the proper motions for most stars \citep{Zerjal24}, they were not able to do so for \sigoristar~since it is saturated ($G=3.4$mag).

In addition to the differences in proper motions, the radial velocity or systemic velocity of  \sigoristar~may also differ to that of the other stars in \sigori. Analysing pre-1990 RV data, \citet{Barbier2000} found $v_{\rm r} = 29.9\pm1.6\rm km~s^{-1}$, similar to that of the other members of \sigori. More recently, \citet{SimonDiaz15} analysed newly obtained spectroscopic data and found $v_{\rm r} = 31.0\pm0.16\rm km~s^{-1}$, consistent with previous estimates. However, more recent spectroscopic measurements \citep{Schaefer16} at the 1.5 m telescope at CTIO using the Chiron spectrometer \citep{Tokovinin13}, found $v_{\rm r}\simeq36.37\pm0.38\rm km~s^{-1}$, whilst when combining their spectroscopic measurements with visual interferometric observations with the CHARA array, they found radial velocities $v_{\rm r} = 37.64\pm0.35 \rm km~s^{-1}$. Although when combining the interferometric data with the spectroscopic data of \citet{SimonDiaz15}, \citet{Schaefer16} found $v_{\rm r} = 31.18\pm0.21 \rm km~s^{-1}$. The analysis by \citet{Schaefer16} also took into account the complex morphology of \sigoristar~that is in fact a triple system, since they accurately determined the orbital parameters of the system using interferometric measurements. They attribute the differences in radial velocities to different methods being used to fit blended lines in the spectroscopic data, as well as differences in wavelength calibrations. In summary, the observed radial velocities of \sigoristar~are similar to the other stars in \sigori, but there could be subtle differences. We also note that the differences in radial velocities is also consistent with observed velocity dispersions found in the Orion region \citep[$\sim$1--3 km s$^{-1}$,][]{VanAltena88,Kim19,Theissen23}.

With both the proper motions of \sigoristar~being significantly different to that of the surrounding cluster, this indicates that it may have a different origin and is only now passing through the cluster as a runaway system. Such a system typically forms within a specific star forming region, before undergoing dynamical interactions with nearby massive stars, resulting in its' ejection \citep{Hoogerwerf2000,Renzo19,CarreteroCastrillo25}. Indeed, there are multiple examples of low mass runaway stars in the ONC that are expected to have formed in proximity to $\theta^1$ Ori C \citep{Mcbride19,Platais20}, whilst other stars have been scattered towards the ONC from other populations in the Orion Complex \citep{Mcbride19}. 

\subsection{An infrared arc}
Whilst there are significant uncertainties on the proper motion of \sigoristar~from \textsc{hipparcos}, due to the complex nature of the system, there are other observable features in the cluster that indicate the direction and magnitude of the proper motion are appropriate.
Indeed, there is an infrared arc observed by WISE and Spitzer associated with \sigoristar~that is approximately aligned with its proper motion vector \citep{Ochsendorf14, Koenig15}. There are two potential explanations for this which we summarise briefly, but both would be sensitive to the proper motion vector.

Massive stars emit a lot of energy into their surroundings both in the form of winds and ionising radiation. If the star is moving sufficiently fast this distorts the wind bubble \citep{Mackey2015} and if supersonic results in the accumulation of material at the shock interface and dusty infrared arcs oriented approximately along the stellar propagation direction \citep[e.g.][]{vanBuren1990, Kaper1997, Mackey2016, Green2019, Green2022}. The on-sky velocity of \sigoristar~is supersonic relative to the ionised gas velocity $\sim10$\,km\,s$^{-1}$ and so a shock would form at the wind interface and accumulate material. \cite{2024A&A...689A.352K} show that the arc is associated with both a gas and dust enhancement. 

The other explanation proposed by \cite{Ochsendorf14} is a dust wave resulting from the equilibrium point between radiation pressure and dust dragged out in a photoevaporative wind from the dark cloud L1630. This works if the wind is relatively weak, which there is some evidence for \citep{Sanz-Forcada2004}. However the wind mass loss rate and velocity is very uncertain. Furthermore at the current stellar velocity over the $\sim10^5\,$yr time-scales for grains to equilibrate \citep{Ochsendorf14} \sigoristar~would have moved over a scale larger than the arc.  

Regardless of which explanation is correct, the brightest part of the infrared arc is exactly along the stellar propagation vector, which is offset by about 40 degrees relative to what would be the photoevaporative flow vector in the dust wave scenario. The morphology of the IR arc hence supports the idea that \sigoristar's propagation vector is genuine. 

\begin{figure*}
\centering
\includegraphics[scale=0.5]{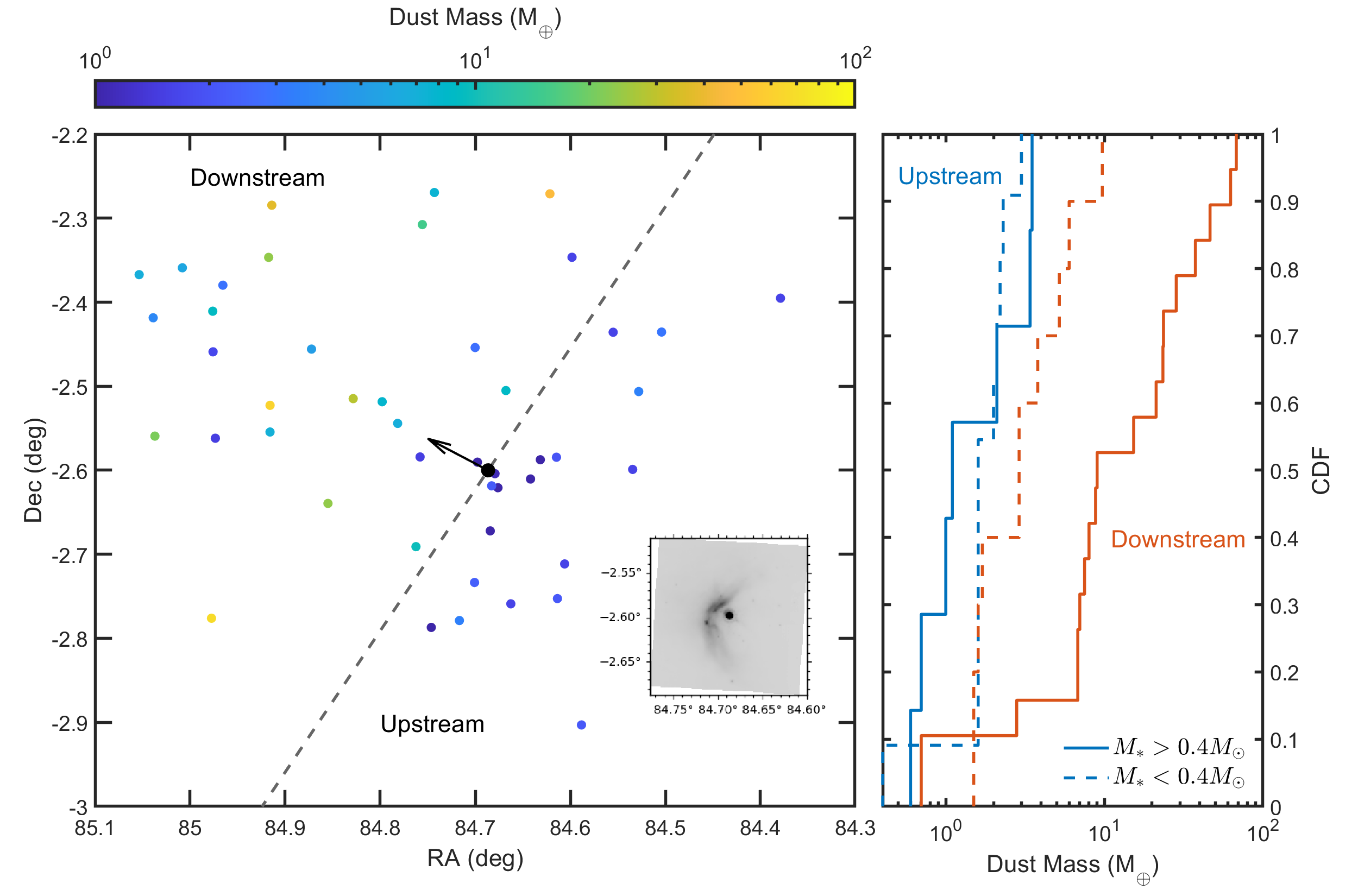}
\vspace{-0.3cm}
\caption{Left-hand panel: A 2D map of \sigori, with the colour showing the observed disc mass \citep{Mauco23}. The black dot represents \sigoristar~with the arrow showing its' proper motion and expected movement over the next 10,000 years. The grey dashed line represents the normal to the trajectory of \sigoristar, splitting the region into a ``downstream'' and an ``upstream''. The inset plot shows a \textit{Spitzer} image of the infrared arc. \citep{Rieke04} Right-hand panel: Cumulative distribution functions for the observed dust disc masses in \sigori, split into an upstream region (blue lines) and a downstream region (red lines). Solid lines show the masses for stars more massive than $0.4 \msun$, with dashed lines showing for those less massive than $0.4 \msun$.}
\label{fig:sig_ori_map}
\end{figure*}
 
\subsection{The origin of \sigoristar}
If \sigoristar~is passing through the cluster at high velocity then this introduces the challenge of determining the formation site. Naive extrapolation of the 2D vector traces towards the south western part of Barnard's loop, where there is (potentially triggered) star formation occurring over the last 1--2\,Myr \citep[][]{Gandolfi2008} and which it would reach in $\sim 1$\,Myr at the current on-sky speed.

However, such thinking is oversimplified. The entire stellar and gas population in the cluster will have changed significantly over the of order $\sim3$ Myr lifetime of \sigori \citep{Grudic2021, Foley2023}. For example, compare the upper right and lower central panels of Figure 1 from \cite{Grudic2021}, which show snapshots from a simulation with a feedback driven feature like Barnard's loop. The sites of star formation in the upper right panel would not be identified as cluster locations or sites of current active star formation 4.4\,Myr later  in the lower central panel. Indeed some of the earlier star forming clumps are associated with voids of gas and stars at the later time. We also do not know at what point in time the event occurred that would have led to the acceleration of \sigoristar, meaning it could have been located relatively close and only recently been made a runaway. Although uncertain, estimates of the ages of the stars in \sigoristar~place them at $\leq1$ Myr \citep{SimonDiaz15,Schaefer16}, much younger than the estimates of \sigori~members \citep[$\sim$3--4 Myr,][]{Sherry08}.

Given the above we hence argue that one shouldn't expect to find a clear separate cluster that \sigoristar~originated in.

\subsection{Disparity in disc masses across \sigori}

Recent observations of \sigori~clearly showed the effects of external photoevaporation on protoplanetary disc masses, where observed discs with projected distances closer to \sigoristar~having smaller dust masses \citep{Mauco23}. They highlight that this is especially the case for more massive stars ($\geq0.4\msun$), where the level of reduction in masses in the inner 0.5 pc of the cluster was over an order of a magnitude compared to those further away. In making their conclusions on how external photoevaporation has affected those discs, they assumed that the cluster was co-moving, i.e. there are minimal proper motion differences between the different stars. As we highlighted above, this may not be the case. By assuming that \sigoristar~is actually moving through the cluster, it is possible to split the region into two, an \textit{upstream} where stars have already interacted with \sigoristar, and a \textit{downstream} where stars are yet to reach their peak of interacting with \sigoristar. Figure \ref{fig:sig_ori_map} shows a 2D representation of the cluster in the left-hand panel, with the dust mass of each observed disc shown by the colour. The black point shows the location of \sigoristar~, with the arrow showing it's proper motion over the next 10,000 years. The inset plot shows a \textit{Spitzer} image of the infrared arc around \sigoristar \citep{Rieke04}. Finally the dashed line represents the normal to the trajectory of \sigoristar, splitting the region in two. It is clear from Fig. \ref{fig:sig_ori_map} that the discs in the upstream region are significantly less massive than those in the downstream region. Indeed the maximum dust disc mass in the upstream region is $\le 4\me$, whilst 80$\%$ of discs in the downstream are more massive than $8\me$. These differences are further highlighted in the right-hand panel of Fig. \ref{fig:sig_ori_map} where the cumulative distribution functions are shown for the upstream (blue) and downstream (red) discs. We split the two distributions into those around low mass stars ($<0.4 \me$, dashed lines), and high mass stars ($\geq0.4\me$, solid lines). For the more massive stars, the disparity between the upstream and downstream dust mass distributions is clear, with discs downstream being at least an order of magnitude more massive than their upstream counterparts. This trend is less clear for discs around low mass stars, but those discs downstream are still a factor few more massive than those upstream.

\section{Simulation Parameters}

With the evidence above, including the observed disc masses, indicating that \sigoristar~may be passing through the cluster, it is useful to understand whether theoretical studies predict such signatures. Recent work explored the effects of a flyby star on the evolution of protoplanetary discs \citep{Coleman25FUV}. They found that discs are significantly truncated when exposed to even short periods of strong irradiation as a massive UV source performs a flyby. Specifically they predicted that discs downstream of the UV source's velocity vector would be more massive and extended than those upstream.

\begin{figure}
\centering
\includegraphics[scale=0.55]{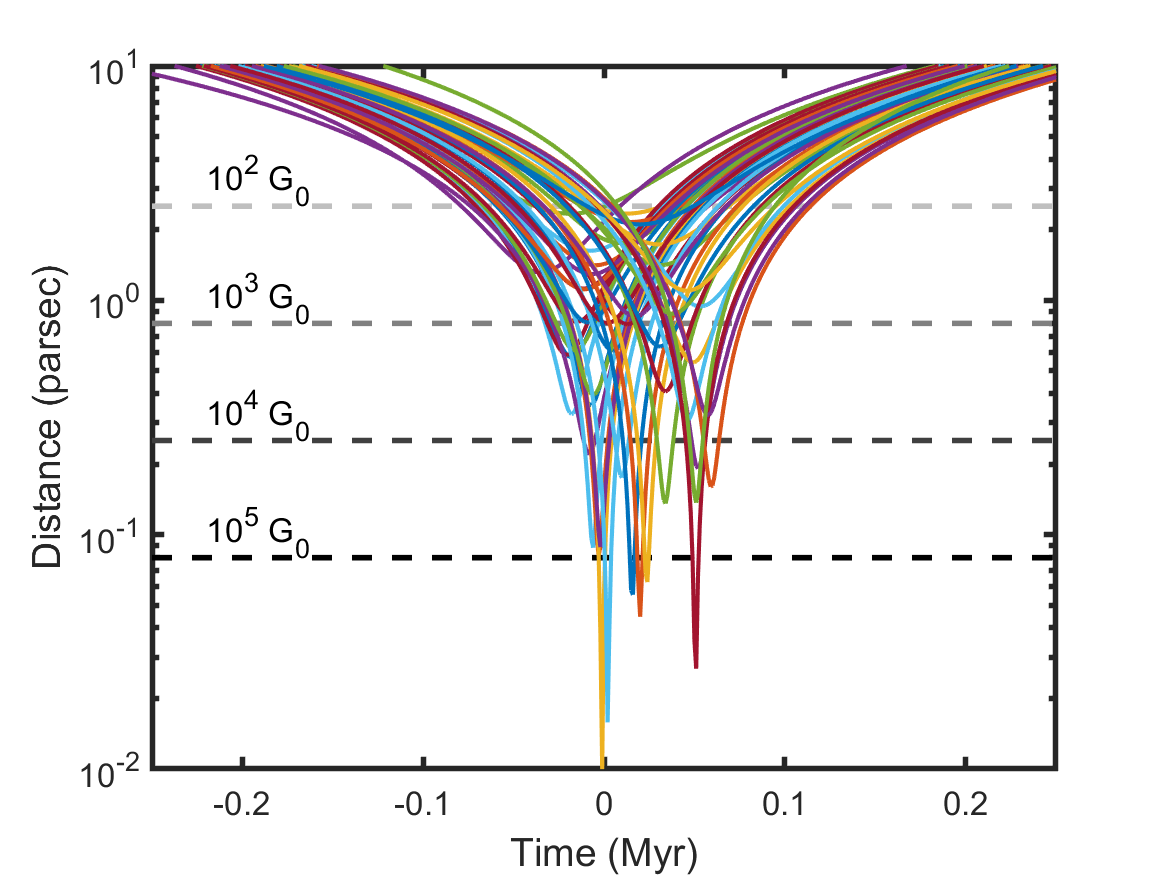}
\caption{Distances of stars in \sigori~to \sigoristar~as it passes through. The dashed horizontal lines show the distances that correspond to different levels of irradiation.}
\label{fig:distances}
\end{figure}

As observations of the discs in \sigori~show evidence for a flyby of \sigoristar, which also has significantly larger proper motions than the other stars in the region, we assume that this is the case and carry out simulations of protoplanetary disc evolution for the discs observed in the cluster. First, using the proper motions of the stars in the cluster \citep{Mauco23,Zerjal24}, we determine the distances of each star to \sigoristar~over time (assuming a distance to the cluster of 387.5pc \citep{Schaefer16}). We assume that the flyby occurs after 2 Myr, and we run the simulations for a total of 5 Myr. Figure \ref{fig:distances} shows the distances from each star to \sigoristar, centred around the peak flyby time of 2 Myr. Horizontal dashed lines indicate varying strengths of the UV field, showing how many of the discs experience extremely high levels of irradiation for short periods of time, compared to their more quiescent evolution when not in the vicinity of \sigoristar. Using these distances, we then calculate the time-evolving UV field strength and apply it to the evolution of the protoplanetary discs.

To evolve the discs we use the disc model that is presented in \citet{Coleman25FUV}, of which we now briefly describe. Protoplanetary discs lose mass through accretion on to the central star, and by photoevaporative winds launched from the surface layers of the disc. To model the gas disc, we use a 1D viscous disc model \citep{Shak} where the evolution of the gas surface density $\Sigma$ is solved through the standard diffusion equation.
\begin{equation}
    \dot{\Sigma}(r)=\dfrac{1}{r}\dfrac{d}{dr}\left[3r^{1/2}\dfrac{d}{dr}\left(\nu\Sigma r^{1/2}\right)\right]-\dot{\Sigma}_{\rm PE}(r)
\end{equation}
where $\nu=\alpha H^2\Omega$ is the disc viscosity with viscous parameter $\alpha$ = $10^{-2.5}$, $H$ the disc scale height, $\Omega$ the Keplerian frequency, and $\dot{\Sigma}_{\rm PE}(r)$ is the rate of change in surface density due to photoevaporative winds.
Following \citet{Coleman25FUV} we include EUV and X-ray internal photoevaporative winds from the central star \citep{Picogna21} as well as winds launched from the outer disc by far ultraviolet (FUV) radiation emanating from nearby massive stars, e.g. O-type stars \citep{Haworth23}.
We assume that the photoevaporative mass loss rate at any radius in the disc is the maximum of the internally and externally driven rates 
\begin{equation}
    \dot{\Sigma}_{\rm PE}(r) ={\rm max}\left(\dot{\Sigma}_{\rm I,X}(r),\dot{\Sigma}_{\rm E,FUV}(r)\right)
\end{equation}
where the subscripts I and E refer to contributions from internal and external photoevaporation.
With the stars in \sigori~varying in stellar mass, this also means they will have different X-ray luminosities. We follow \citet{Flaischlen21} that found a linear relation between the stellar mass and the X-ray luminosity.
\begin{equation}
    \log{L_X} = 30.58 + 2.08\log{\left(\frac{M_*}{\msun}\right)}. 
\end{equation}
For the external photoevaporative winds, we use the \textsc{fried} grid \citep{Haworth23}, that provides mass loss rates for discs irradiated by FUV radiation as a function of the star/disc/FUV parameters. Following the method outlined in \citet{Coleman25FUV} we use the fiducial \textsc{fried} grid of $f_{\rm PAH}=1$ (an interstellar medium, ISM-like PAH-to-dust-ratio), and with the assumption that grain growth has occurred in the outer disc, depleting it and the wind of small grains which reduces the extinction in the wind and increases the mass loss rate compared to when dust is still ISM-like. This combination of parameters results in PAH-to-gas abundances comparable to our limited observational constraints on that value \citep{Vicente13,Schroetter25}. For the FUV field strength, we take the calculated values based on the distances to \sigoristar~as shown in Fig. \ref{fig:distances}. We refer the reader to Section 2 of \citet{Coleman25FUV} for details on the full implementation of the model.

\section{Explaining the distributions in disc properties}

\begin{figure}
\centering
\includegraphics[scale=0.4]{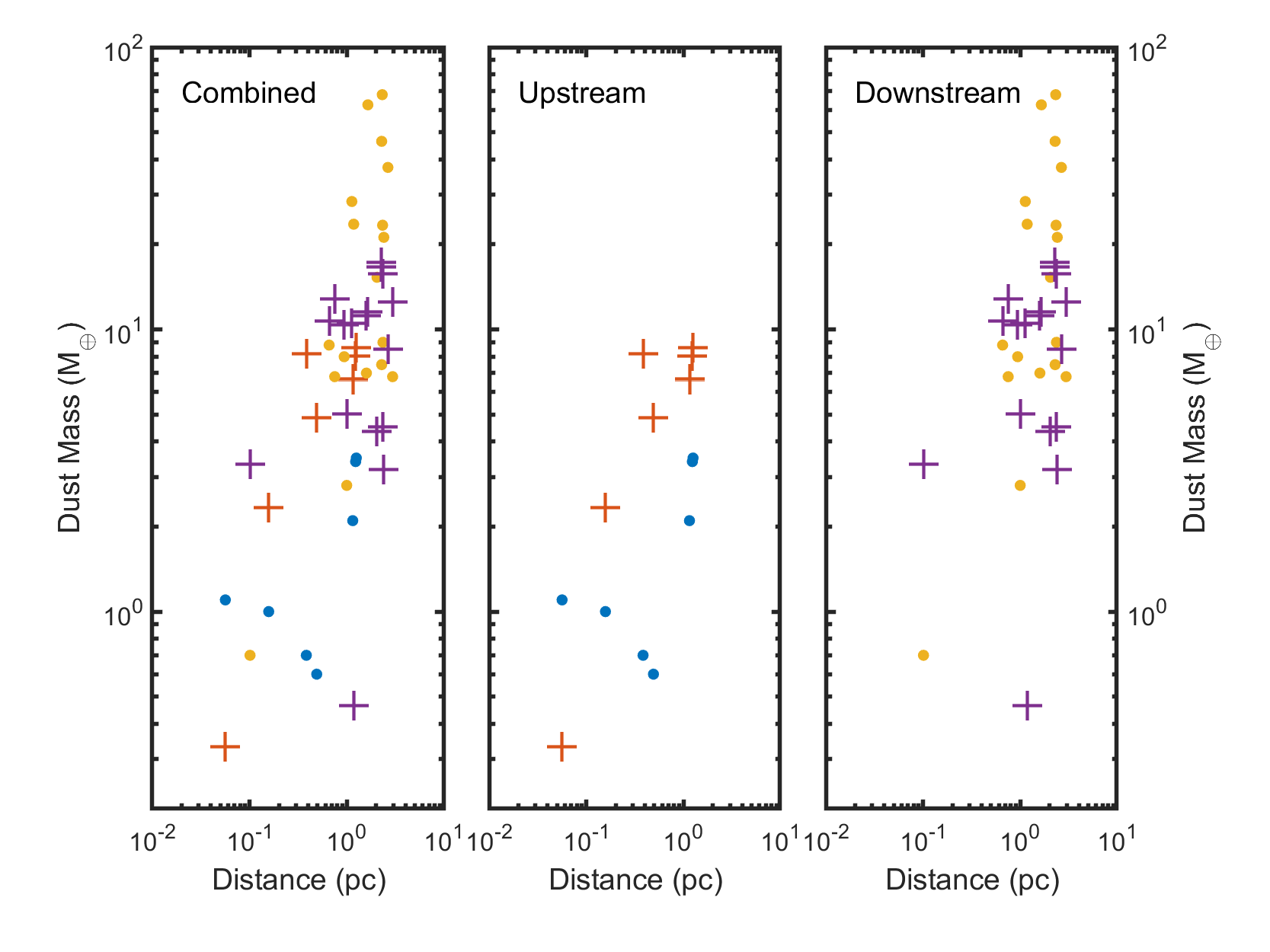}
\caption{Dust disc masses versus projected distance from the current location of \sigoristar~for discs that are downstream (right-hand panel), upstream (middle panel), and combined (left-hand panel). Observations are denoted by dots, and simulated discs are denoted by plus signs. Only those discs around more massive stars ($M_*\geq0.4\msun$) are plotted.}
\label{fig:mass_comparison}
\end{figure}

For each of the discs, we evolved them for 5 Myr with the flyby of \sigoristar~occurring after 2 Myr. The main observable in regards to the discs is the dust disc mass as was shown in Fig. \ref{fig:sig_ori_map}. In Fig. \ref{fig:mass_comparison} we show the dust masses of the discs after 2 Myr, i.e. the current time, as a function of the distance to \sigoristar~ at that point in time. The right-hand panel shows the discs that are downstream of the trajectory of \sigoristar, the central panel shows those discs that are upstream, whilst the left-hand panel shows a combination of the two, similar to how such observations are typically presented. The observed discs are shown as points, whilst the simulation results are shown by plus symbols. For the simulated discs, we take the dust mass as being equal to 1/500th of the gas disc mass, assuming that the initial dust-to-gas ratio was ISM-like, i.e. 1/100th, whilst also applying an additional reduction of 80\% to account for dust conversion into pebbles and planetesimals over the lifetime of the disc, similar to that used in planet formation models \citep[e.g.][]{Brugger20,Coleman24,Coleman24FFP}.
Comparing the downstream to the upstream regions, it is clear that both the observed and simulated discs consistently have higher dust masses in the downstream region compared to the upstream region when comparing discs of similar projected distances to \sigoristar. This is the result of the increased total amount of radiation received by the discs upstream of \sigoristar.

Comparing the observations to the simulations in Fig. \ref{fig:mass_comparison}, we can see there is good agreement with simulations being able to match the observations to within factor few. The simulations do however underestimate the dust masses in the downstream region, which could indicate that the observed discs are more efficient at retaining dust than the simulations, possibly through dust traps \citep{Garate2024}. Additionally, for those discs that are upstream of \sigoristar, the simulations overestimate the remaining dust mass for most discs, implying that the inner discs in the observed systems are more depleted than that predicted by typical disc evolution models.
On the whole, these differences between the individually observed and simulated discs could arise from a myriad of possibilities, concerning both the observations and simulations. These include: the 3D nature of the cluster is not well known; the levels of turbulence as well as the initial disc masses and sizes may differ from disc to disc; more complex effects such as planets and dust traps affect the disc evolution; as well as other uncertainties in measurements and dust-to-gas conversions. Ultimately, these possibilities will affect the observed dust masses and sizes, either by determining the initial quantity, or by slightly affecting their evolution.
Interestingly, when looking at the combined disc masses in the left-hand panel of Fig. \ref{fig:mass_comparison}, it is clear that both the observations and simulations provide a more continuous distribution of dust masses as a function of distance, than if the upstream/downstream regions are explored independently, similar to that stated in previous works \citep{Mauco23}.

\begin{figure}
\centering
\includegraphics[scale=0.4]{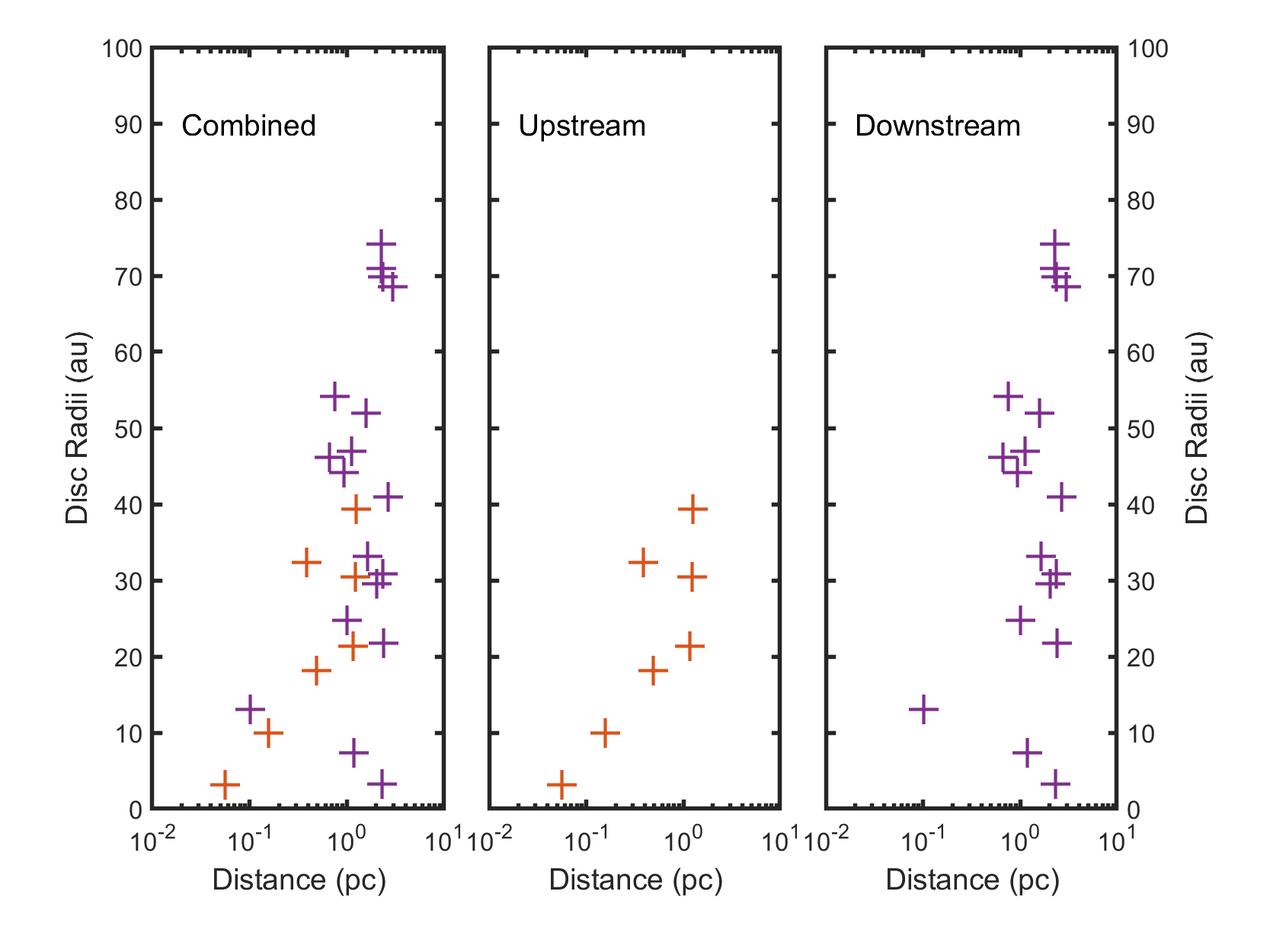}
\caption{Simulated disc radii versus projected distance from \sigoristar~for discs that are downstream (right-hand panel), upstream (middle panel), and combined (left-hand panel). Only those simulated discs around more massive stars ($M_*\geq0.4\msun$) are plotted.}
\label{fig:radii_comparison}
\end{figure}

Whilst the simulations are able to match the trends seen in the observations of disc mass, they also predict that the differences in disc radii either side of the moving \sigoristar~will be clearly observable \citep{Coleman25FUV}. Indeed, Fig. \ref{fig:radii_comparison} shows the disc radii as a function of distance to \sigoristar~for the simulated discs for those downstream (right-hand panel), and upstream (central panel) of the system, as well as the combined population (left-hand panel). The differences between the disc radii downstream and upstream of \sigoristar~is clear to see, with those discs upstream all being smaller then $40 \au$, whilst over half of the downstream discs are all greater than $40 \au$, with some being as extended as $75 \au$. When looking at the combined population, it is also clear that the differences is not a projected distance effect since there is significant overlap in distances between those discs upstream and downstream, with those downstream containing many discs that are significantly larger than those upstream.

\section{Discussion and Conclusions}

In this work we explored the idea that \sigoristar~is actually a runaway system that is performing a flyby of the \sigori~cluster of stars. With multiple indications that the system is indeed moving quickly through the cluster, we then ran disc evolutionary models to explain features observed in disc population studies of the region. We summarise our main results below:

(1) The observed proper motions of \sigoristar~indicate that it is substantially different to that of the other stars in \sigori. Additionally, the morphology of an IR arc observed in \textit{WISE} and \textit{Spitzer} images is consistent with the apparent trajectory of \sigoristar.

(2) By splitting the discs observed in \citet{Mauco23} into two regions, upstream and downstream of \sigoristar, we show that those discs that are upstream, have significantly lower disc masses than those that are downstream. This signature is even more prominent when only considering discs around high mass stars ($M_* \geq0.4\msun$).

(3) Using the observed high proper motion of \sigoristar, we applied a time-evolving FUV field to our disc evolution models, and showed that they could match the observed signatures when comparing disc masses between the upstream and downstream populations.

(4) In addition to being able to match the differences in disc masses between the upstream and downstream population, the disc models also predict that there will be a disparity in disc radii between the two regions. The discs in the upstream region are all expected to be compact, $\le40\au$, whilst the majority of discs in the downstream region are more extended, with some reaching $75\au$.

In summary, our results show that there are multiple indications that \sigoristar~is a runway system and is performing a flyby of \sigori, depleting the discs residing there of gas through powerful photoevaporative winds. Further observations of \sigori, especially those that confirm the proper motions of \sigoristar, should confirm the nature of the system. If this is confirmed, this would show that the dynamics of massive stars is extremely important for studies of many aspects of protoplanetary disc evolution, as those clusters are not static with the conditions that they evolve in constantly changing over time. Additionally, by understanding the dynamical history of \sigori, this can make it an ideal region to explore protoplanetary disc evolution models.

\section*{Data Availability}
The data underlying this article will be shared on reasonable request to the corresponding author.

\section*{Acknowledgements}
The authors thank the anonymous referee for their useful and insightful comments that improved the quality of the paper.
GALC acknowledges funding from the UKRI/STFC grant ST/X000931/1.
TJH acknowledges UKRI guaranteed funding for a Horizon Europe ERC consolidator grant (EP/Y024710/1) and a Royal Society Dorothy Hodgkin Fellowship. 
JSK acknowledges NASA’s Nexus for Exoplanet System Science (NExSS) research coordination network sponsored by NASA’s Science Mission Directorate and project “Alien Earths” funded under Agreement No. 80NSSC21K0593.
This research utilised Queen Mary's Apocrita HPC facility, supported by QMUL Research-IT (http://doi.org/10.5281/zenodo.438045).
\bibliographystyle{mnras}
\bibliography{references}{}

\label{lastpage}
\end{document}